\documentclass[3p,times,twocolumn]{elsarticle}

\usepackage{ecrc}

\volume{00}

\firstpage{1}

\journalname{Nuclear Physics B Proceedings Supplement}

\runauth{}

\jid{nuphbp}

\jnltitlelogo{Nuclear Physics B Proceedings Supplement}

\usepackage{amssymb}
\usepackage{ifthen}
\usepackage{subfig}%
\usepackage{xspace}%
\usepackage{multirow}%
\usepackage{amsmath,amssymb}% for advanced math typesetting
\usepackage{pennames-pazo}% particle names: converted from pennames for mathpazo

\usepackage[figuresright]{rotating}

\usepackage{ptdr-definitions}
\newcommand\HBB{\ensuremath{\PH\to\mathrm{\cPqb\cPaqb}}\xspace}
\newcommand\mH{\ensuremath{\mathrm{m}_{\mathrm{H}}}\xspace}
\newcommand{\Vvar}{\ensuremath{{\mathrm{V}}}\xspace}

\newcommand\WlnH{\ensuremath{\PW(\ell\cPgn)\PH}\xspace}
\newcommand\WtnH{\ensuremath{\PW(\Pgt\cPgn)\PH}\xspace}

\newcommand\ZnnH{\ensuremath{\cPZ(\cPgn\cPgn)\PH}\xspace}
\newcommand\ZllH{\ensuremath{\cPZ(\ell\ell)\PH}\xspace}

\newcommand\ptV {\ensuremath{\pt(\Vvar)}\xspace}

\newcommand\dphiVH {\ensuremath{\Delta\phi(\Vvar,\PH)}\xspace}

\newcommand\ZtoLL {\ensuremath{\cPZ\to\ell\ell}}
\newcommand\ZtoNN {\ensuremath{\cPZ\to\cPgn\cPagn}}
\newcommand\dphiMJ {\ensuremath{\Delta\phi(\MET, \mathrm{jet})}}

\newcommand\Nal {\ensuremath{N_{\mathrm{a}\ell}}}
\newcommand\ptjj   {\ensuremath{{\pt}(\mathrm{jj})}}

\newcommand\Mjj     {\ensuremath{m(\mathrm{jj})}}
\newcommand\Naj {\ensuremath{N_{\mathrm{aj}}}}

\newcommand\mtau       {\ensuremath{\tau}\xspace}
\newcommand\dphiMtkM {\ensuremath{\Delta\phi(\MET,{\MET}_{\text{tracks}})}}
\newcommand\dPhiMETlep {\ensuremath{\Delta\phi(\MET,\ell)}}

\newcommand{\BDT}{BDT\xspace}

\newcommand{\Mbb}{\ensuremath{\mathrm{M}_{\bbbar}}}

\newcommand{\muBDT}{\ensuremath{\mu = 1.09 {}_{-0.21}^{+0.24}}}

\def\ttH  {\ensuremath{\mathrm{\ttbar H}}}

\def\HBB {\ensuremath{\mathrm{H}\to\bbbar}}

\def\ttBB  {\ensuremath{\ttbar \bbbar}}

\def\ZnnH     {\ensuremath{\mathrm{Z}(\cPgn\cPgn)\mathrm{H}}}
\def\ZllH     {\ensuremath{\mathrm{Z}(\ell\ell)\mathrm{H}}}

\def\WlnH     {\ensuremath{\mathrm{W}(\ell\cPgn)\mathrm{H}}}

\def\WtnH     {\ensuremath{\mathrm{W}(\tau\cPgn)\mathrm{H}}}

\def\ZtoNN    {\ensuremath{\mathrm{Z}\to\cPgn\bar{\cPgn}}}
\def\ZtoLL    {\ensuremath{\mathrm{Z}\to\ell\ell}}

\begin{document}

\begin{frontmatter}

%% Title, authors and addresses

%% use the tnoteref command within \title for footnotes;
%% use the tnotetext command for the associated footnote;
%% use the fnref command within \author or \address for footnotes;
%% use the fntext command for the associated footnote;
%% use the corref command within \author for corresponding author footnotes;
%% use the cortext command for the associated footnote;
%% use the ead command for the email address,
%% and the form \ead[url] for the home page:
%%
%% \title{Title\tnoteref{label1}}
%% \tnotetext[label1]{}
%% \author{Name\corref{cor1}\fnref{label2}}
%% \ead{email address}
%% \ead[url]{home page}
%% \fntext[label2]{}
%% \cortext[cor1]{}
%% \address{Address\fnref{label3}}
%% \fntext[label3]{}

\dochead{}
%% Use \dochead if there is an article header, e.g. \dochead{Short communication}

\title{Search for the Higgs boson in the $\mathrm{b\bar{b}}$ decay channel using the CMS detector}

%% use optional labels to link authors explicitly to addresses:
%% \author[label1,label2]{<author name>}
%% \address[label1]{<address>}
%% \address[label2]{<address>}

\author{Caterina Vernieri for the CMS Collaboration}

\address{Scuola Normale Superiore, Piazza dei Cavalieri, 7, 56126 Pisa}

\begin{abstract}
A search for the standard model Higgs boson in the \bbbar decay channel has been carried out with the CMS detector at the LHC collider. The searched production modes are the associated VH production, the VBF production and the production in association with top quark pairs (ttH). The analyses are based on pp collision data collected at centre-of-mass energies of 7 and 8 TeV, corresponding to integrated luminosities of 5 fb$^{-1}$ and 20 fb$^{-1}$, respectively. The strategy and results of the searches are reported.
%% Text of abstract
\end{abstract}

\begin{keyword}
%% keywords here, in the form: keyword \sep keyword
Higgs \sep b-quark \sep b-tagging \sep jet \sep regression \sep energy resolution
%% MSC codes here, in the form: \MSC code \sep code
%% or \MSC[2008] code \sep code (2000 is the default)
\end{keyword}
\end{frontmatter}

%%
%% Start line numbering here if you want
%%
% \linenumbers

%% main text
%\section{}
%\label{}

%%%>  Introduction <%%%
\section{Motivations}
At a mass ($\mH$) of 125 GeV the standard model (SM) Higgs boson decay mode into a bottom quark-antiquark pair (\bbbar) dominates the total width~(${\sim}58\%$)~\cite{LHCHiggsCrossSectionWorkingGroup:2011ti}. While the decay of the Higgs boson to vector bosons has been observed in different channels (ZZ, $\gamma\gamma$,WW)~\cite{ZZ,Hgg,WW}, the direct couplings of the Higgs boson to fermions, and in particular to down-type quarks, remains to be firmly established~\cite{Nature-14,ATLAS-CONF-2014-009}. 

The current measurements constrain indirectly the couplings to the up-type top quark, since the dominant Higgs production mechanism is gluon fusion induced by top-quark loop. The measurement of the \HBB\ decay represents a direct test of whether the observed boson interacts as expected with the quark sector and provides the unique final test of the direct coupling of the Higgs boson to down-type quarks, an essential aspect of the nature of the newly discovered boson.  To date, the most precise constraints on the couplings to down-type quarks are provided by the CMS experiment~\cite{Chatrchyan:2008aa}, exploiting the results from the VH(\bbbar) search~\cite{PRD-89-012003}.

%%Evidence for the direct decays to fermions coupling of the Higgs boson has been recently reported by CMS combining the search of Higgs decays to bottom quarks and $\tau$ leptons, leading to the first evidence for the direct coupling of the 125 GeV Higgs boson to down-type fermions, with an observed significance of 3.8 ~$\sigma$, when 4.4 being expected~\cite{Nature-14}.
%%The direct measurement of the Higgs boson decays to fermions, as recently reported by CMS in the study of Higgs decays to bottom quarks and $\tau$ leptons has lead to the first strong evidence for the direct coupling of the 125 GeV Higgs boson to down-type fermions, with an observed significance of 3.8 ~$\sigma$, when 4.4 are expected~\cite{Nature-14}.\\
%

\section{Challenges}
Despite the largest expected branching fraction, the \HBB\ final state is quite more challenging to measure compared to the cleaner signatures provided by the decay in ZZ and then in four leptons or in two photons that have led to the Higgs boson discovery at LHC. 

Besides the poorer invariant mass resolution, the \bbbar final state is also characterized by smaller signal over background ratio. The dominant production mechanism is gluon fusion ($\sim 87\%$ at LHC), but when paired with the \bbbar decay mode, the resulting irreducible background from QCD production of b quarks is overwhelming, roughly 7 order of magnitude larger than the signal. The signal topology of other production mechanisms is exploited in order to increase the sensitivity to the \HBB\ signal. The most sensitive channel at LHC is the search for the SM decay \HBB\ in events where the Higgs boson is produced in association with a W or Z boson decaying leptonically.  After event selections, VH(\bbbar) signal events are one order of magnitude more than the H$\rightarrow 4\ell$ events, 
however they are spread over an interval of mass that is ten times larger and with a signal over background a factor 40 smaller.

%\begin{table}[h!]
%
%\centering{
%\scalebox{0.75}{
%\begin{tabular}{ccc}
%\hline
%       &H$\rightarrow 4\ell$ &       \HBB\ \\
%\hline
%BR &	 0.013\%	&58\% \\
%mass resolution 	&1\%	&10\%\\
%signal efficiency	& 30\%	&	1.3\% \\
%S/B 	&2 &	0.05 \\
%
%\hline
%
%\end{tabular}
%}}
%\caption{Comparison in terms of signal efficiency and S/B between the H$\rightarrow 4\ell$ and \HBB\ decay modes. For the latter the VH production mechanism is assumed. }
%\label{tab:comparisonVH1}
%\end{table}

%In Tab.~\ref{tab:comparisonVH1} a comparison with the four leptons decay mode is reported, to justify the small sensitivity of the \HBB\ decay mode. After the proper event selections, VH(\bbbar) signal events are three order of magnitude more than the H$\rightarrow 4\ell$ events, however they are spread over an interval of mass that is ten times larger and with a signal over background a factor 40 smaller.
%Despite the largest branching fraction the \HBB\ decay mode results in a smaller sensitivity with respect to the four leptons decay mode. 

\section{b-jet Identification}

The identification or ``b-tagging" of jets resulting from the fragmentation and hadronization of bottom-quark plays a fundamental role in reducing the otherwise overwhelming background to these signal signatures from processes involving jets from gluons (g), light-flavor quarks (u, d, s), and c-quark fragmentation.

A useful property of B hadrons in this respect is their lifetime, with $c\tau\sim$ 500~$\mu$m. Therefore a B hadron with a momentum of 50 GeV will fly on average almost half a centimeter before decaying. This translates into the fact that secondary decay particles will have a sizable impact parameter (IP) with respect to the B hadron production point. 
%half a centimeter before decaying ($L\sim \gamma c\tau $) 
%The impact parameter $d\sim L \sin(\alpha) \sim \gamma c\tau \alpha\sim c\tau$ is boost invariant, where $\alpha$ is the average opening angle of the decay products.

The B hadrons are much more massive than anything they decay into, thus the decay products will have a few GeV of momentum in the B rest of frame. This effect is particularly evident for leptons from B decays, which have order of a GeV of \pt relative to the B flight direction, while leptons in generic jets (from decays in flight of $\pi$'s or K's) tend to be more closely aligned with the jet. Another peculiar property of the B hadron decay is the relatively high rate of lepton production from leptonic decays~($\sim$~35\%); indeed, the presence of leptons is a good signature of the presence of B hadrons in a jet. 
%, which reduces the effects of multiple scattering and allows these impact parameters to be measured with good resolution. Another peculiar ..
A variety of reconstructed objects -- tracks, vertices and identified leptons -- are used to build observables, which are then combined into a single discriminating variable, which separates b from light-flavored jets.
The main observables used as input to the b-tagging algorithms are related to the B hadron lifetime and the presence of a secondary vertex.

The IP of a track with respect to the primary vertex can be used to distinguish decay products of a \cPqb\ hadron from prompt tracks. The IP is calculated in three dimensions, taking advantage of the excellent resolution of the pixel detector.  For transverse momenta above a few GeV, since the multiple scattering is less relevant, the resolution on the two-dimensional IP is independent of $\eta$ and approaches 30~$\mu$m~\cite{trk-11-001}.

The presence of a secondary decay vertex and kinematic variables associated with this vertex can be used to discriminate between b and non-b jets. These variables include the flight distance and direction, i.e. the vector between primary and secondary vertex, and various properties of the system of associated secondary tracks such as the multiplicity, the invariant mass or the energy.  

The CSV b-tagging algorithm~\cite{BTV12001} is used to identify b-jets in the CMS searches reported here. This algorithm combines in an efficient way the information about track IP and secondary vertices, providing discrimination also when no secondary vertices are found.

The efficiency to tag b jets and the rate of misidentification of non-b jets depend on the requirement on the discriminant, and on the transverse momentum and pseudorapidity of the jets.  Several working points for the CSV output discriminant are used in the analyses.  The loose (CSVL), medium (CSVM), and tight (CSVT) operating points are defined as the CSV values such that the a misidentification probability for light-parton jets is close to 10\%, 1\%, and 0.1\%, respectively, at jet-\pt of about 50 GeV and $\eta\sim0$.
For the CSVT requirement the efficiencies to tag b quarks and c quarks are approximately 50\% and 6\% respectively. The corresponding efficiencies for CSVM are approximately 65\% and 15\%, while for CSVL are 81\% and 32\%.

\section{Specific Corrections for $b$-jet Transverse Momentum}\label{sec:regression}
%the ATLAS experiment is about 12\%~\cite{ATLAS-CONF-2013-079}, to be compared with
The invariant mass resolution plays a fundamental role to improve the \HBB\ signal discrimination against the non-peaking background and to achieve a better separation of the Higgs boson from the Z boson peak. \\
The jet energy calibration in CMS is performed as a function of the jet-\pt and $\eta$, and taking into account
the pile-up activity of the event. The calibration is quite accurate, however it is performed on a QCD dijet sample composed mainly by gluon initiated jets and does not take into account the additional details of the jet reconstruction. 

Typical values for the jet energy resolution in CMS are about 15\% for \pt$=$ 30 GeV~\cite{JME-10-014}. The presence of neutrinos in more than 35\% of the B hadron decay chains results in a lower response for the b-jet, with respect to the light quark/gluon induced jets used in the standard calibration. 

%Since jet angles are measured with a relative better resolution than the jet-\pt,  
In CMS, the searches for the \HBB\ (both VH and VBF) have developed a specific strategy to correct the b-jet energy to improve the invariant mass resolution of the reconstructed Higgs boson. By applying multivariate regression techniques similar to those used by the CDF experiment~\cite{1107.3026} the method attempts to recalibrate the jet energy to the true b-jet energy. 
%For example,
%it is expected that jets with high charged-hadron fraction will have a higher response than those with low. 
Thanks to the particle-flow jet reconstruction~\cite{pft-09-001}, one can access various properties of each jet and adjust the calibration accordingly. The regression is essentially a multi-dimensional calibration to the particle level - including neutrinos - which exploits the main b-jet properties.  A specialized BDT~\cite{tmva} is trained on simulated signal events and provides for each jet a correction factor that improves both the b-jet energy scale and its resolution. 

Inputs are chosen among variables that are correlated with the b-quark energy and well measured. They include detailed jet structure information about tracks and jet constituents, which differs from light flavor quarks/gluons jets. Information from B-hadron decays on the reconstructed secondary vertices and the soft lepton (SL) from semi-leptonic decay are used, providing an independent estimate of the b quark \pt.  Also the information carried by the variables related to the \MET vector is exploited for the channel where no real missing transverse energy is expected (\ZtoLL, VBF), acting as a kinematic constraint for the momentum balance in the transverse plane.

%This procedure especially addresses the problem of semi-leptonic b decays. 
The most discriminating variables across all modes are kinematic, and this
is due to the fact that most of the power of the regression derives
from the neutrinos involved in the semi-leptonic B decays. 
%This effect is illustrated
%in Fig.~\ref{fig:Semilep}, which shows the dijet mass before and after
%regression for the case where the b jet contains a muon, and when it
%does not, for \ZllH\ signal events. 
%\begin{figure}[h!]
%  \begin{center}
%      \includegraphics[width=0.48\textwidth]{Figures/ZllH_DiJetPt_Oct5_Data_mBB_Sig125_HMassRegression_NoSemi.pdf}
%    \includegraphics[width=0.48\textwidth]{Figures/ZllH_DiJetPt_Oct5_Data_mBB_Sig125_HMassRegression_Semi.pdf}
%    \caption{Comparison of the reconstructed dijet invariant mass for
%    Higgs candidates \ZllH\ signal events before and after the
%    regression.  Separate plots are shown for the case where the b jet
%    does not (left) or does (right) contain a muon from semi-leptonic
%    B decay.}
%    \label{fig:Semilep}
%  \end{center}
%\end{figure}

%However, also for \ZllH\ channel the improvement is larger in case of a
%semileptonic b decay.
%Therefore soft lepton variables are included
%in the regression across all modes.

In Tab.~\ref{tab:listInputsReg} the variables used as input to the regression in the VH search are listed. 
Regression performance depend on the phase space used to train the BDT in a non negligible way, so separate trainings are performed for each channel with very similar sets of inputs to deal with the different kinematic properties. 
%In Appendix~\ref{app:appendix_regressionVBF} more details on the input and phase space optimizations are reported for the VBF channel.

%show comparisons of data and MC for several control regions, demonstrating
%that backgrounds are not biased by the regression technique.

\begin{table}[h!]
\centering{
\scalebox{0.85}{
\begin{tabular}{l}
\hline
Variable \\
\hline
  jet-\pt  before calibration \\
  jet-\pt after the default calibration \\
  jet \et after the default calibration \\
  jet transverse mass after the default calibration \\
  uncertainty on the JEC\\
  % \item $\eta$ -- pseudorapidity of the jet;
  transverse momentum of the leading track in the jet  \\
  secondary vertex decay length  \\
  error on jet secondary vertex decay length $dL$ \\
  jet secondary vertex mass \\
  jet secondary vertex \pt \\
 
 %\item Chf -- fraction of jet constituents that are charged;
 %\item Nch -- number of jet constituents that are charged;
  total number of jet constituents \\
  relative transverse momentum of SL candidate to the jet-\pt \\
  transverse momentum of SL candidate in the jet \\
  distance in $\eta-\phi$ of SL candidate with respect to the jet axis \\
% \item $\rho 25$ -- energy density calculated within $\left | \eta \right | < 2.5$;
  event total transverse missing energy (\MET) \\
  azimuthal opening between \MET and the jet directions \\
%   jet-\pt (after the default calibration)\\
%   jet $\eta$\\
%   jet mass \\
%   jet neutral-hadron energy fraction \\
%   jet photon energy fraction \\
%   jet secondary vertex mass \\
%   error on jet secondary vertex decay length $dL$ \\
%   azimuthal opening between the missing energy and the jet directions $d\phi(MET,\pt)$ \\
%   event total transverse missing energy (MET) \\
%   transverse momentum of the leading track in the jet \\
%   total number of jet constituents \\
%   transverse momentum of soft lepton candidate in the jet \\
%   relative transverse momentum of soft lepton candidate to the jet-\pt\\
   % which makes them easier to identify and separate from lepton sources in generic jets \\
\hline

\end{tabular}
}}
\caption{Set of variables used as input to the regression for the VH channel. }
\label{tab:listInputsReg}
\end{table}
 
\begin{figure}[tbhp]
 \begin{center}
    \includegraphics[width=0.4\textwidth]{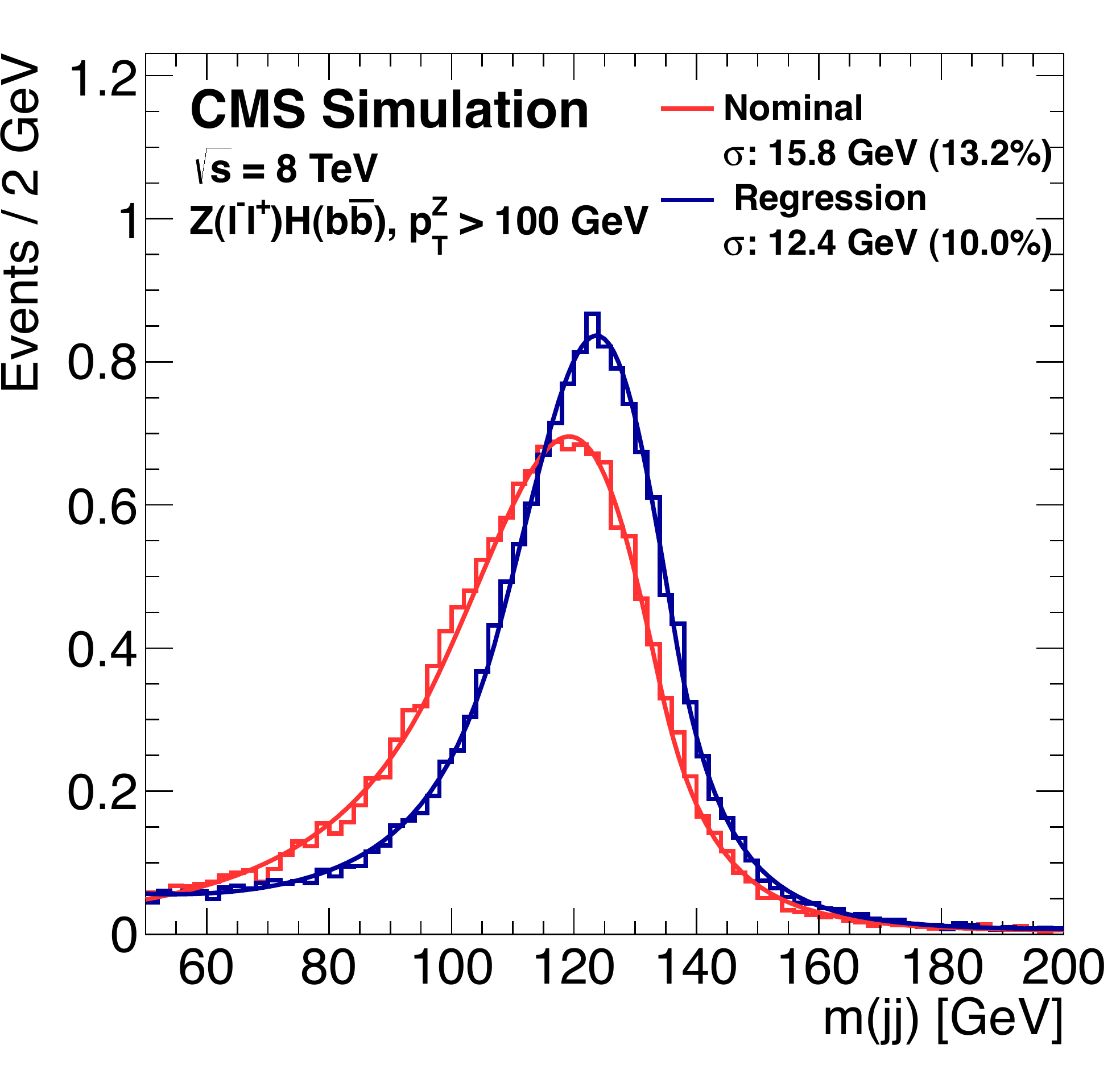}
    \caption{Dijet invariant mass distribution for simulated samples of \ZllH\ events before (red) and after (blue) the energy correction from the
regression procedure is applied. 
%Right. Comparison of the invariant mass of the two jets with the highest b-tag CSV values from VBF signal events, with default jets, and \pt-regressed. A Gaussian function is fit to the core of the  distribution and a third order Bernstein polynomial to model the tails.
    }
    \label{fig:regression_VV_VH}
  \end{center}
\end{figure}

\begin{figure}[h!]
\begin{center}
\includegraphics[trim = 0mm 10mm 0mm 0mm, width=0.4\textwidth]{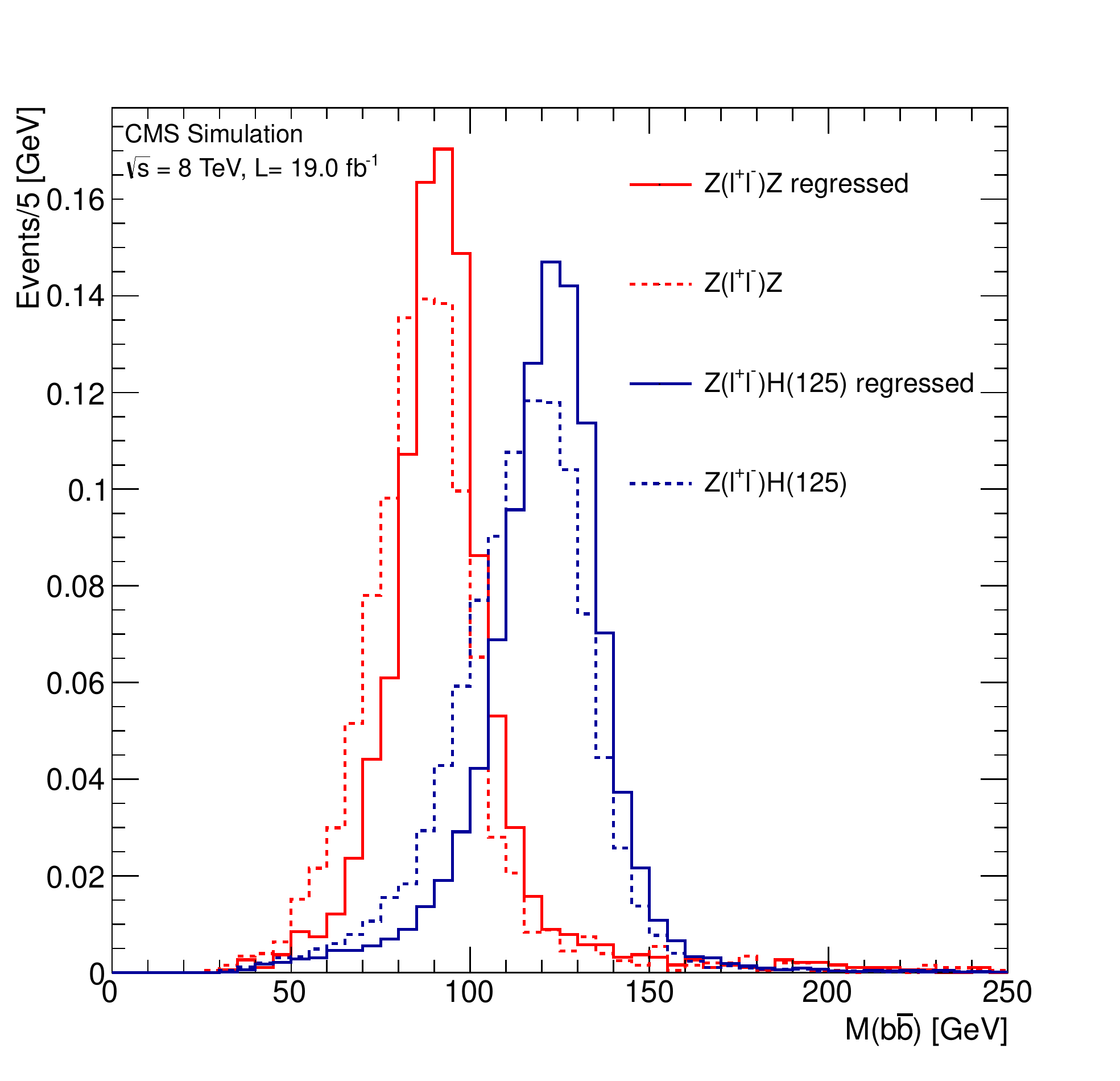}
\end{center}
\caption{Mass difference between ZH and ZZ simulated processed before and after the regression is applied.}
\label{fig:MassDiff}
\end{figure}

The average improvement on the mass resolution, measured on simulated signal samples, when the corrected jet energies are used is $\approx$15-25\%, resulting in an increase in the analysis sensitivity of 10--20\%, depending on the specific phase space. This improvement is shown in Fig.~\ref{fig:regression_VV_VH} for simulated samples of $\ZllH$ events. A better separation of the VZ/VH signals is also achieved by applying this corrections, as reported in Fig.~\ref{fig:MassDiff} for $\ZllH$ and Z$(\ell\ell)$Z simulated events.

The validation of the regression technique in data has been performed studying the \pt balance in Z($\ell\ell$)$+\bbbar$ events and the reconstructed top-quark mass distribution in \ttbar-enriched samples targeting the lepton+jets final state.
%In the following section we report the dijet balance study and in sec~\ref{sec:vzMeas} the VZ$\to\bbbar$ cross section measurement is reported.  

\section{VH(\bbbar) Searches}\label{sec:vhhasd}
%\textsc{\large{Higgs Strahlung}},\hspace{3mm} i.e. the {associated production} of a Higgs boson and a W or Z boson is characterized by an even smaller production cross section than vector boson fusion. The main contribution 
The most sensitive channel for the search for the SM decay \HBB\ at the LHC is the associate production with a W or Z boson, when the Higgs boson recoils
with large momentum transverse to the beam line ($\pt > 100\GeV$). 

%In their combined search for the SM Higgs boson~\cite{PhysRevD.88.052014}, the CDF and D0 collaborations at the Tevatron \Pp\Pap\ collider have reported evidence for an excess of events in the 115--140\GeV mass range, consistent with the mass of the Higgs boson observed at the LHC. In that search, the sensitivity below a mass of 130GeV is dominated by the channels in which the Higgs boson is produced in association with a weak vector boson and decaying to \bbbar~\cite{PhysRevLett.109.071804}. The combined local significance of this excess is reported to be 3.0 ~$\sigma$ at $\mH=125$~GeV, while the expected local significance is 1.9 ~$\sigma$.  \\

%At the LHC the VH channel is the most sensitive channel for the . The search for the SM decay \HBB\ is performed in a phase space selected by requiring a significant boost of the \pt of the vector boson, \ptV, or of the Higgs boson ($\pt > 100\GeV$) 

%The ATLAS experiment using data samples corresponding to an integrated luminosity of $4.7$~fb$^{-1}$ at $\sqrt{s}=7~\TeV$ reported no significant excess, for the Higgs boson mass hypothesis of 125 GeV, a 95\% CL$_S$$_{s}$ upper limit of 1.4 times the SM expectation is set on the cross section times branching ratio for \HBB\ ~\cite{ATLAS-CONF-2013-079}. 
The CMS experiment reported an excess of events above the expected background, the first 2$\sigma$ indication of the \HBB\ decay at the LHC. The result combines the analysis of the data samples corresponding to integrated luminosities of up to $5.1$~fb$^{-1}$ at $\sqrt{s}=7~\TeV$~\cite{Chatrchyan:2013lba} with the analysis of the full 8\TeV data sample corresponding to a luminosity of up to $18.9$~fb$^{-1}$.

%The results presented here
%combine the analysis of the 7\TeV data sample
%in~\cite{Chatrchyan:2013lba} with an updated analysis of an 8\TeV
%data sample corresponding to $18.9$~fb$^{-1}$,
%that includes the $5.3$~fb$^{-1}$ data sample in~\cite{Chatrchyan:2013lba}.

%The search is performed in data samples corresponding to integrated
%luminosities of $5.1$~fb$^{-1}$ at $\sqrt{s}=7\TeV$ and up to
%$18.9$~fb$^{-1}$ at $\sqrt{s}=8\TeV$. %~\cite{Chatrchyan:2012ww} = this is the first Hbb 7 TeV publication

%$\mathrm{W}(\mu\nu)\mathrm{H}$, $\mathrm{W}(e\nu)\mathrm{H}$,
%$\mathrm{W}(\tau\nu)\mathrm{H}$, $\mathrm{Z}(\mu\mu)\mathrm{H}$,
%$\mathrm{Z}(ee)\mathrm{H}$, and $\mathrm{Z}(\nu\nu)\mathrm{H}$

%$\mathrm{W}(\mu\nu)\mathrm{H}$ and $\mathrm{W}(e\nu)\mathrm{H}$

The search analyzes six channels separately for the V decay mode: $\mathrm{W}(\mu\nu)\mathrm{H}$,
   $\mathrm{W}(\mathrm{e}\nu)\mathrm{H}$,
$\mathrm{W}(\tau\nu)\mathrm{H}$, $\mathrm{Z}(\mu\mu)\mathrm{H}$, $\mathrm{Z}(\mathrm{ee})\mathrm{H}$, and
   $\mathrm{Z}(\nu\nu)\mathrm{H}$, all with the Higgs boson decaying to \bbbar. For each channel, different
\ptV\ boost regions are selected.  Because of different signal and background
content, each \ptV\ region has different sensitivity and
the analysis is performed separately in each region, resulting in a total of 14 channels.

%\footnote{Only the 8~\TeV data are
%included and only taus with 1-prong
%hadronic decays are explicitly considered, reconstructed using the Hadron Plus Strips (HPS) algorithm~\cite{Tau:2012}}

The presence of a vector boson decaying leptonically in the final state highly suppresses the QCD background, while also providing an efficient trigger path. The requirement that the Higgs boson is produced with a large boost provides several additional advantages: it exploits the harder \pt spectrum of the signal, thus resulting in a further reduction of the large backgrounds from W and Z boson production in association with jets; it helps in reducing the large background from top-quark production in the signal channels including neutrinos; it enables the accessibility of the \ZtoNN\ channel at trigger level via large \MET; and it generally improves the mass resolution of the reconstructed Higgs candidates.

The background processes to VH production with \HBB\ are the production of vector
bosons in association with one or more jets (V+jets), \ttbar production,
single-top-quark production, diboson production (VV),
and QCD multijet production. Except for dibosons, these processes have
production cross sections that are several orders of magnitude larger
than Higgs boson production.

\begin{table}[tbp]
\caption{Selection criteria that define the signal region. 
The values listed for kinematic variables are in units of \GeV, and
for angles in units of radians.}

\label{tab:BDTsel}
\begin{center}
\scalebox{0.70}{
\begin{tabular}{lcccc} \hline
Variable                     & \WlnH                           &     \WtnH             & \ZllH                 & \ZnnH                               \\\hline
\ptV                         & $[100-130]$   &
$[>120]$             & $[50-100]$    & $[100-130]$  \\ 
 &  $[130-180]$    &
             &  $[>100]$   &  $[130-170]$   \\ 

 &  $[>180]$    &
          &    &  $[>170]$  \\ \hline

$m_{\ell\ell}$               & --                              &      --               & $[75-105]$            & --                                  \\
$\pt(\mathrm{j}_1)$                   & $>30$                           &     $>30$             & $>20$                 & $>60$                               \\
$\pt(\mathrm{j}_2)$                   & $>30$                           &     $>30$             & $>20$                 & $>30$                               \\
\ptjj                        & $>100$                          &
$>120$            & --                    & $>$100, 130, 130             \\
\Mjj                         & $<250$                          &
$<250$             & 40-250, $<$250   & $<250$                              \\
\MET                         & $>45$                           &   $>80$               & --                    &  see \ptV\\
$p_{\mathrm{T}}(\mtau)$               & --                              &      $>40$            & --                    & --                                  \\
$p_{\mathrm{T}}(\mathrm {track})$               & --                              &     $>20$             & --                    & --                                  \\
CSV$_{\mathrm{max}}$         & $>0.40$                         &
$>0.40$          & $>$0.50, 0.244    & $>0.679$                            \\
CSV$_{\mathrm{min}}$         & $>0.40$                         &      $>0.40$          & $>0.244$              & $>0.244$                            \\
%CSV$^{loose}_{\mathrm{min}}$ & -- ($<0.40$)                   &       --              & -- ($<0.244$)         & -- ($<0.244$)                       \\
\Naj                         & --                              &
--              & --                    & $<$2, --, --                      \\
\Nal                         & $=0$                            &      $=0$             & --                    & $=0$                                \\
\dphiVH                      & --                              &      --               & --                    & $>2.0$                              \\
\dphiMJ                      & --                              &
--               & --                    & $>$0.7, 0.7, 0.5          \\
\dphiMtkM                    & --                              &      --               & --                    & $<0.5$                              \\
\MET significance            & --                              &
--               & --                    & $>$3, --, --                    \\
\dPhiMETlep                  & $<\pi/2$                        &          --             & --                    & --                                  \\
%Tightened Lepton Iso.        & $< 0.075$ (--, --)              &    --                 & --                    & --                                  \\
%BDT                         & full distribution    & full distribution     & full distribution         &                  \\           
\hline
\end{tabular}
}
\end{center}
\end{table}

Key point of the analysis strategy is the use of the {regression}, to improve the invariant mass resolution as described in sec.~\ref{sec:regression}. The reconstructed \HBB\ mass resolution achieved is about 10\%.

Control regions in data are selected to adjust the event yields from simulation for the main background processes in order to estimate their contribution in the boosted phase space defined by the signal region.  

\begin{figure}[h!]
  \begin{center}
   \includegraphics[width=0.4\textwidth]{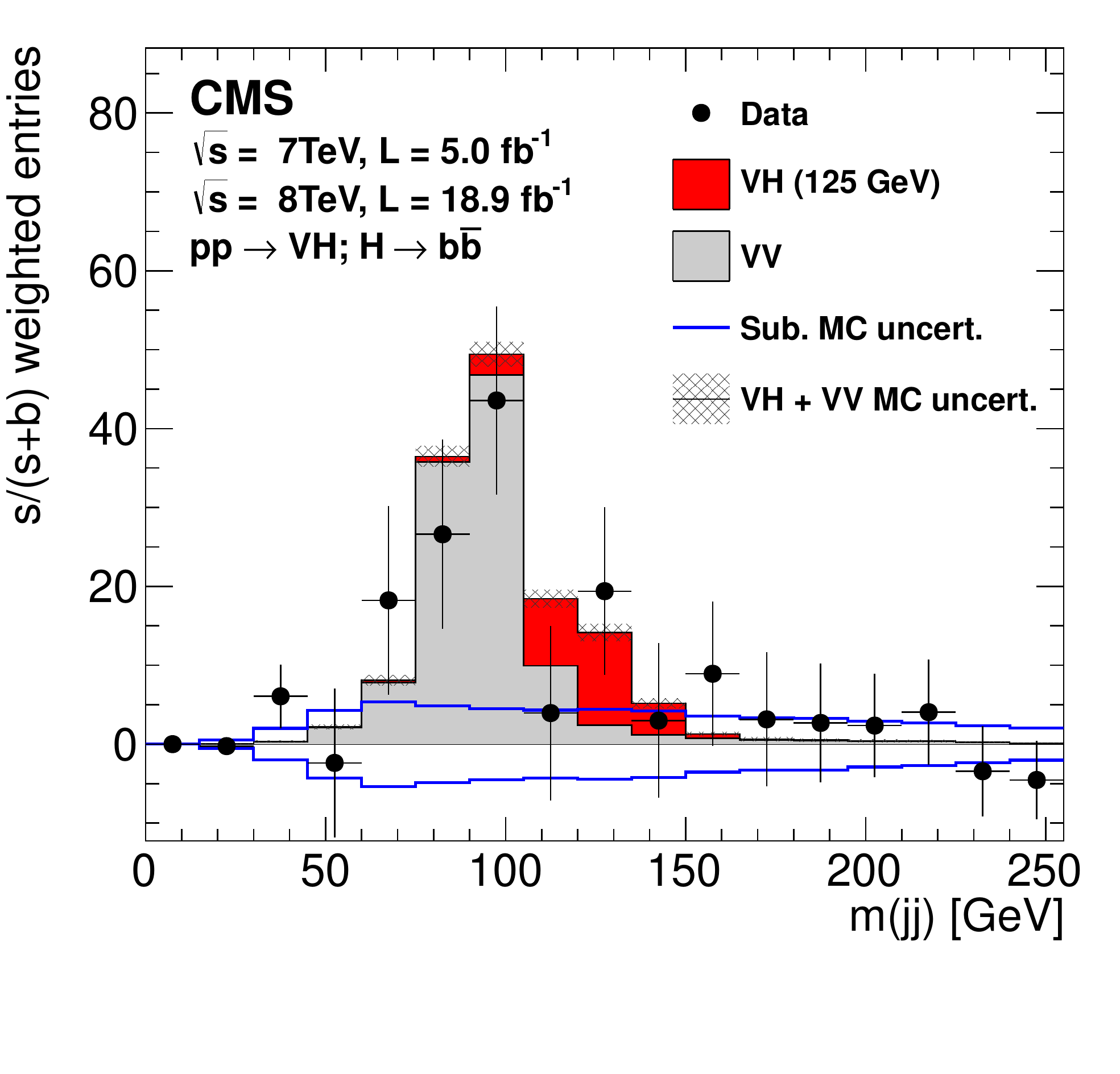}
  \caption{ Dijet invariant mass distribution combined for all
channels after the non resonant subtraction. }
    \label{fig:MJJ-combined}
  \end{center}
\end{figure}

%Appropriate control regions are identified in data and used to
%validate the proper simulation modeling of the distributions used as input to the BDT
%discriminants, and to correct the simulation normalization for several of the most important background processes:
%production of W and Z bosons in association with jets (light- and heavy-flavor) and \ttbar\ production~\cite{PRD-89-012003} and typically include inverted b-tag selections and mass sidebands.
%Separate scale factors are applied for each background
%process in the different channels.
%
The signal region is defined by the requirements listed in Tab.~\ref{tab:BDTsel}. Fig.~\ref{fig:MJJ-combined} shows the \Mbb\ distribution after the non resonant backgrounds subtraction. The VH signal is 
visible with a yield compatible with the SM expectation and a significance of 1.1$\sigma$.

To better separate signal from background under
different Higgs boson mass hypotheses, an event BDT discriminant is trained separately at
each mass value using simulated samples for signal and all background
processes. The training of this BDT is performed with all events in the signal region.
Among the most discriminant
variables for all channels are the dijet invariant mass distribution, the number of additional jets, the value of CSV for the
Higgs boson daughter
with the second largest CSV value, and the angular
distance between Higgs boson daughters. 
By making use of correlations between discriminating variables in signal and background events, the BDT yields a new variable that allows to discriminate the signal more effectively than with the use of the \Mbb\ information only. 

\begin{figure}[htbp]
  \begin{center}
     \includegraphics[width=0.33\textwidth]{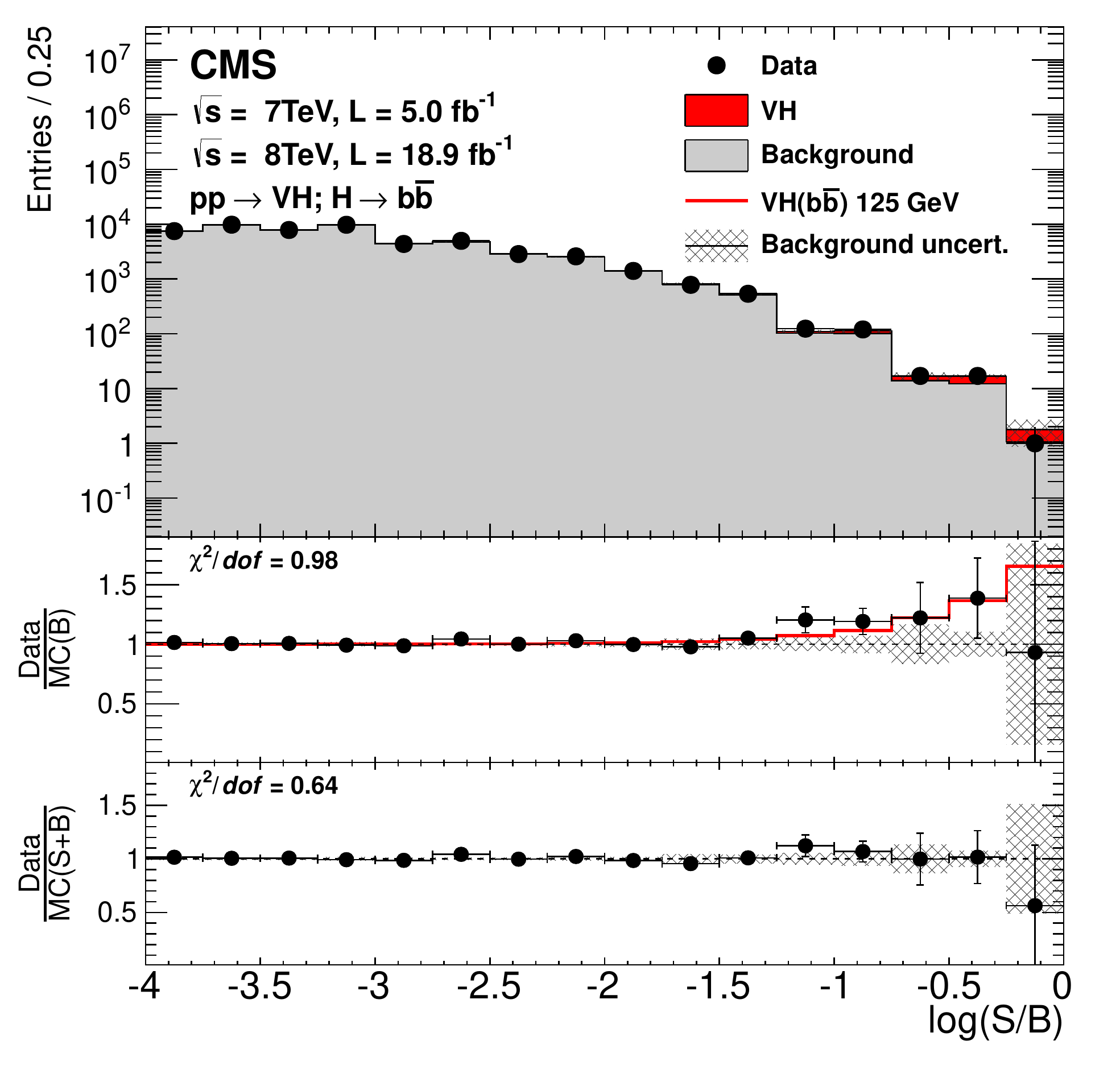}
   \caption{Events are sorted in bins of similar expected signal-to-background as given by the BDT discriminant output value.}
    \label{fig:BDT_S_over_B_all}
  \end{center}
\end{figure}

Results are obtained from combined signal and
background binned likelihood fits to the shape of the output distribution of the BDT
discriminants trained separately for each channel. In total 14 BDT distributions are considered, one for each channel and each boost category. In the simultaneous fit to all channels, in all boost regions, the BDT shape and
normalization for signal and for each background component 
are allowed to vary within the systematic and statistical uncertainties.  Figure~\ref{fig:BDT_S_over_B_all}
combines the BDT outputs of all channels where the events
are sorted in bins of similar expected signal-to-background
ratio, as given by the value of the output of their corresponding BDT
discriminant. The
observed excess of events in the bins with the largest
signal-to-background ratio is consistent with what is expected from
the production of the SM Higgs boson.

%The systematic uncertainties that affect the results presented in this
%article are listed in
%Tab.~\ref{tab:syst}.

The combined effect of the systematic uncertainties results in a reduction of 15\% on
the expected significance of an observation when the Higgs boson is
present in the data at the predicted SM rate.

\begin{figure}[h!]
  \begin{center}
 \includegraphics[width=0.33\textwidth]{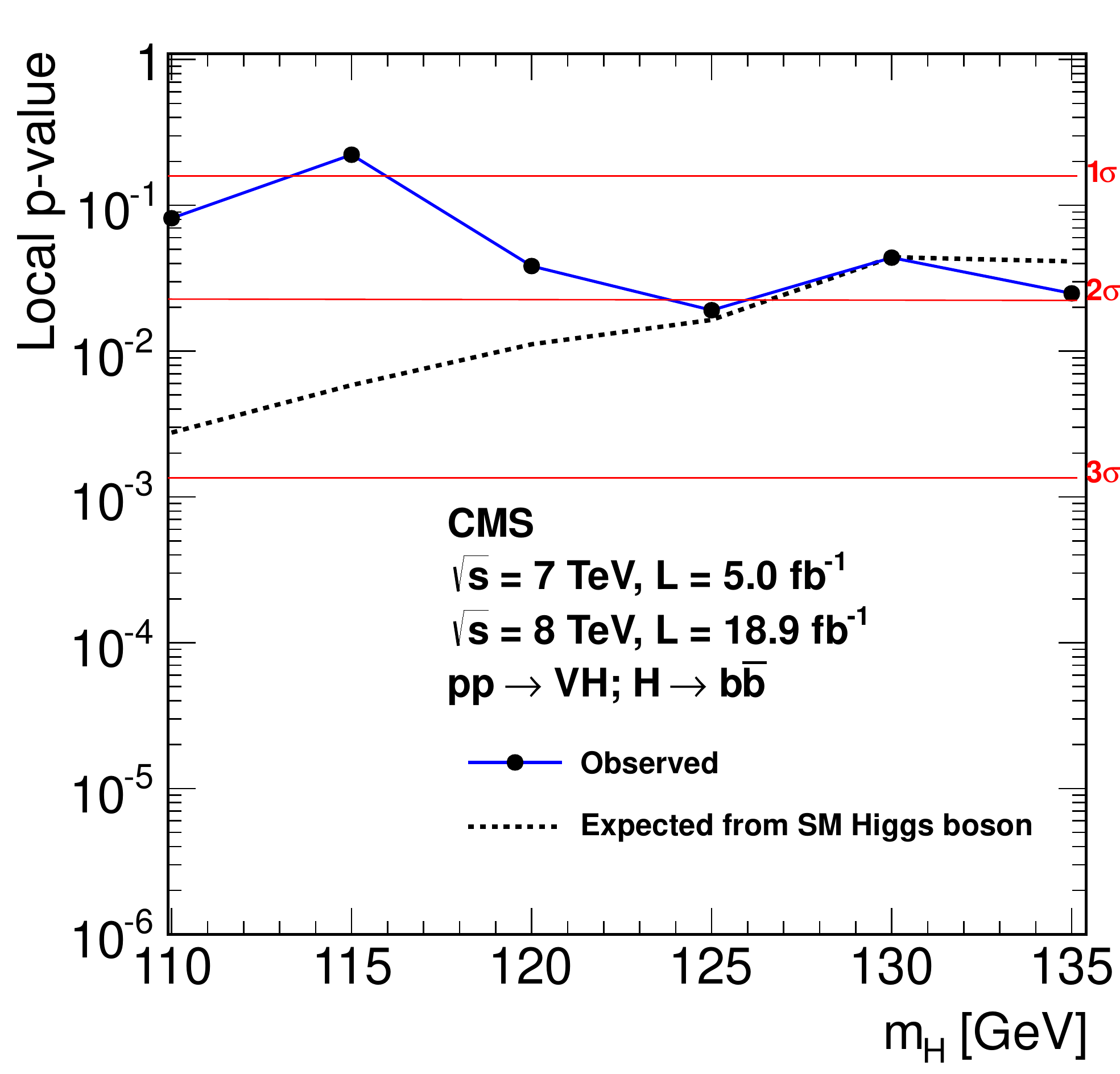}
\caption{Local p-values and corresponding significance (measured in $\sigma$) for the background-only hypothesis.}
    \label{fig:Limits}
  \end{center}
\end{figure}

The analysis sets a 95\% confidence level limit of 1.89 times the SM production for an Higgs boson mass of 125\GeV. The expected limit in absence of Higgs boson production is 0.95. 
%The probability (p-value) to observe data as discrepant as observed under the background-only hypothesis is shown in as a function of the assumed. 
For \mH $=125$\GeV, the excess of observed events corresponds to a local significance of 2.1~$\sigma$ away from the background-only hypothesis, see Fig.~\ref{fig:Limits}.
This is consistent with the 2.1~$\sigma$ expected when assuming the SM prediction for Higgs boson production. The signal strength corresponding to this excess, relative to the expected cross section for the SM Higgs boson, is ${1.0}_{-0.5}^{+0.5}$~\cite{PRD-89-012003}.

%
%In Fig.~\ref{fig:MJJ-combined} the dijet invariant mass distribution is reported
%for the combination of all channels, in all
%boost regions, in the
%combined 7 and 8\TeV data, using the event selection for the \Mjj\
%cross-check analysis~\cite{PRD-89-012003}. 
%Figure~\ref{fig:MJJ-combined} also shows the same weighted
%dijet invariant mass 
%distribution with all backgrounds, except diboson production, subtracted. 

%For the \Mjj\ analysis, a
%fit to the dijet invariant mass distribution results in a measured Higgs boson signal
%strength, relative to that predicted by the standard model, of
%$\mu = 0.8\pm 0.7$, with a local significance of 1.1 standard deviations
%with respect to the background-only hypothesis.
%For a Higgs boson of mass 125\GeV, the expected and observed 95\% CL upper limits
%on the production cross section, relative to the standard model
%prediction, are 1.4 and 2.0, respectively. 
The measured coupling $\kappa_\mathrm{b}^2 = \left.\Gamma_{\mathrm {b\bar b}}\middle/\Gamma^{\mathrm{SM}}_{\mathrm{b\bar b}}\right. $, which quantifies the ratio of the measured Higgs boson partial width into \bbbar relative to the SM value is consistent with the expectations from the SM, within uncertainties.\\
This result is the first 2$\sigma$ indication of the \HBB\ decay at the LHC. The sensitivity of this search - expected significance - is the highest thus far, as reported in Tab.~\ref{tab:comparisonVH}.

% with a local significance of 2.1 ~$\sigma$, which is consistent with the expectation from the production of the SM Higgs boson. 
\begin{table}[h!]
\centering{
\scalebox{0.8}{
\begin{tabular}{lcc}
\hline
Experiment       & $\mu= \sigma/\sigma_{\textnormal{SM}}$&       Expected significance \\
\hline
CDF &	2.5  $\pm$  1.0	&1.3 $\sigma $ ~\cite{Aaltonen:2012if}  \\
D0	&1.2  $\pm$ 1.1	&1.5 $\sigma $  ~\cite{Abazov:2012qya}\\
D0+CDF	& 1.95 $\pm$  0.75&	1.9 $\sigma $~\cite{PhysRevLett.109.071804} \\
ATLAS	&0.2  $\pm$  0.9&	1.6 $\sigma $ ~\cite{ATLAS-CONF-2013-079}\\
CMS	 &1.0 $\pm$ 0.5 & 2.1 $\sigma $ ~\cite{PRD-89-012003}\\
\hline

\end{tabular}
}}
\caption{Signal strength and expected significance for the VH searches at Tevatron and LHC colliders.}
\label{tab:comparisonVH}
\end{table}

The combination of the CMS result with the evidence of the Higgs boson decay to $\tau$ leptons~\cite{doi:10.1007/JHEP052014104} results in the strong evidence (with an observed significance of 3.8~$\sigma$, when 4.4 are expected) for the direct coupling of the 125 GeV Higgs boson to down-type fermions adding precious information to the comprehension of the newly discovered boson. 

ZH, as WH, is dominated by quark-initiated subprocesses (q$\bar{\mathrm{q}}$ZH), but there is also a large gluon-initiated contribution, ggZH, whose contribution to the total cross section is not negligible, 8\%. NNLO QCD corrections to qqZH are included in the VH result reported in~\cite{PRD-89-012003} both as inclusive and as \pt-differential corrections, while NLO corrections to ggZH are applied as a flat over \pt factor~\cite{Altenkamp:2012sx}. The \pt spectrum of the gluon-initiated contribution to associated production has became available recently~\cite{PhysRevD.89.013013}. It is different from the dominant quark-initiated contribution and it peaks at \pt(H)$\sim$150 GeV.  In the estimation of Higgs boson signals in the boosted regime the resulting \pt differential correction due to ggZH contribution impacts up to 30\% the cross section at the highest \pt category. Folding in the corrected ZH \pt spectrum, combining with WH as is, overall VH theory prediction scales up by 10\%, which translates into a roughly 10\% decrease (increase) in $\mu$ (sensitivity).

\subsection{Diboson Cross Section Measurement}
\label{sec:vzMeas}
The Z$\rightarrow \bbbar$ is the purest \bbbar resonance candle, which allows to test b-tagging performance and b-jet energy specific corrections. 
The production cross section for the VZ process, where
$\mathrm {Z}\to\bbbar$, is only few times larger than the VH
production cross section. Given the nearly identical final state this process provides a
benchmark against which the Higgs boson search strategy can be tested. The Z peak, whose resolution is improved by using the aforementioned b-jet specific energy corrections, is also used to measure the consistency of the diboson rate with the expectation
from the SM. This important background to the VH Higgs production with H$\to\bbbar$ is measured in the relevant phase space for the Higgs boson search. 
%Given the different \ptV dependence in VZ and VH processes, as shown in Fig.~\ref{fig:pTdiff} for ZH/ZZ, low p$_T$ regions are more sensitive to the VZ production and higher \pt regions to VH production. 

%Thus the diboson process represents a standard candle to validate the Higgs boson search strategy. 

%
%
%\begin{table}[h!]
%\centering    
%\hspace{-5mm}                     
%\begin{tabular}{c  c c c }
%\hline
%&\multicolumn{3}{c}{$\sigma \cdot $BR at $\sqrt{s} =$8 TeV [pb]}\\
%\hline     
%\bbbar  & \WToLN\ & \ZToLL\  & \ZToNN\   \\
%\hline               
%{Z} &   1.13 & 0.08 & 0.24\\
%{H}  & 0.13 & 0.01 & 0.04 \\
%\hline 
%   
%\end{tabular}
%\caption{Theoretical cross-sections in $pp$ collisions at $\sqrt{s} = 8$~TeV at next-to-leading order (NLO) for the Z boson produced in the mass region $60<\MZ<120$~GeV~\cite{Campbell:2010ff}}
%\label{tab:XSEC}
%\end{table}

%\begin{figure}[h!]
%%\hspace{-7mm}
%\centering
%\hspace{3mm}
%\includegraphics[width=0.35\textwidth]{Figures/ZZ_prod.pdf}
%%\includegraphics[width=0.27\textwidth]{Figures/PAS__ZH125_.pdf}
%\caption{Tree level s-channel Feynman diagram for VZ production at LHC. }
%\label{fig:diag}
%\end{figure}
%

%\noindent
The study of VZ diboson production in proton-proton collisions is per se interesting because it provides an important test of the electroweak sector in the SM.

Using the \BDT\ analysis, the VZ process is observed with a statistical significance of
6.3 ~$\sigma$ (5.9~$\sigma$ expected). This corresponds
to a signal strength relative to the SM of \muBDT. In Fig.~\ref{fig:MJJ-combined} the VZ signal is clearly
visible with a yield compatible with the SM expectation and a significance of 4.1$\sigma$.
%The cross-check analysis (\mbb) yields a significance of 4.1 standard
%deviations (4.6~$\sigma$ expected), which corresponds to \muMbb.
%
All cross sections extracted for individual channels
provide compatible values with each other and the SM expectation. 
%To extract the WZ and ZZ 
%cross-section a simultaneous fit floating both contributions
%independently is performed.
%The best-fit is found to be at \muWZ\ 
%and \muZZ. The resulting measured cross section for WZ and ZZ are measured in a fiducial region for a V transverse momentum above $\ptV > 100GeV$ and found to be compatible with the MCFM calculations, as reported in. 

The 6.3$\sigma$ first observation of the VZ(\bbbar) SM candle at an hadron collides results in a very strong validation of the full analysis strategy designed to reconstruct the \HBB\ decay mode in the VH search at CMS~\cite{Chatrchyan:2014aqa}. 

%\XSWZ\, for the Z boson produced in the mass region $60<\MZ<120$~GeV
%to be compared to the theoretical value of
%\theoryXSWZ\
% calculated with MCFM using the MSTW08 PDF. 
%The ZZ cross section is measured to be
%\XSZZ\, for the Z bosons produced in the mass region $60<\MZ<120$~GeV
%and can be compared to the theoretical value of
%\theoryXSZZ\
% calculated with MCFM
%using the MSTW08 PDF. 
%The uncertainties of both
%theory calculations
%include
%the uncertainties
%on PDF$+ \alpha_s$ and those
%originating from the
%variation of the QCD scales.

\section{ VBF Production Mechanism }

In the search for \HBB\  signals the CMS experiment has also exploited the VBF production mechanism~\cite{CMS-PAS-13-11}, characterized by a very particular event topology.  
In the VBF process a valence quark of each one of the colliding protons radiates a W or Z boson that subsequently interacts or ``fuses". Each valence quark carries on average 1/6 of the proton energy and in the radiation of the weak boson a t-channel four-momentum with Q$^2\sim \mathrm{m}^2_\mathrm{W}$ is exchanged. In this way the two valence quarks are typically scattered away from the beam line and inside the detector acceptance.  

The prominent signature of VBF is therefore the presence of two energetic hadronic jets, roughly in the forward and backward direction with respect to the proton beam line. As a result, in the case of a VBF Higgs boson production, the signal final state features are a central b-quark pair (from the Higgs boson decay) and a light- quark pair (u,d-type) from each of the colliding protons, in the forward and backward regions.
Another important property of signal events is that, the Higgs boson being produced in a VBF process, no color is exchanged in the production.
Thus, in the most probable color evolution of these events, the tagging VBF light-quark jets connect to the proton remnants in the $\pm z$ beam line directions, while
the two b-quarks connect between themselves, as the decay products of the color neutral Higgs. In this way very little additional QCD radiation and hadronic activity is expected in the space outside the color-connected regions, in particular in the whole rapidity interval (rapidity gap) between the two tagging jets, with the exception of the Higgs boson decay products.

Although the predicted cross section production is larger than the VH one, this search results to be less sensitive because the all hadronic final state is quite challenging to trigger on and the QCD multi jet background dominates the selected data sample. Other relevant backgrounds arise from: hadronic decays of Z or W bosons in association with additional  jets, hadronic or semi-leptonic decays of top-pairs, and hadronic decays of single-top productions. 

\begin{figure}[h!]
  \begin{center}
    \includegraphics[width=0.43\textwidth]{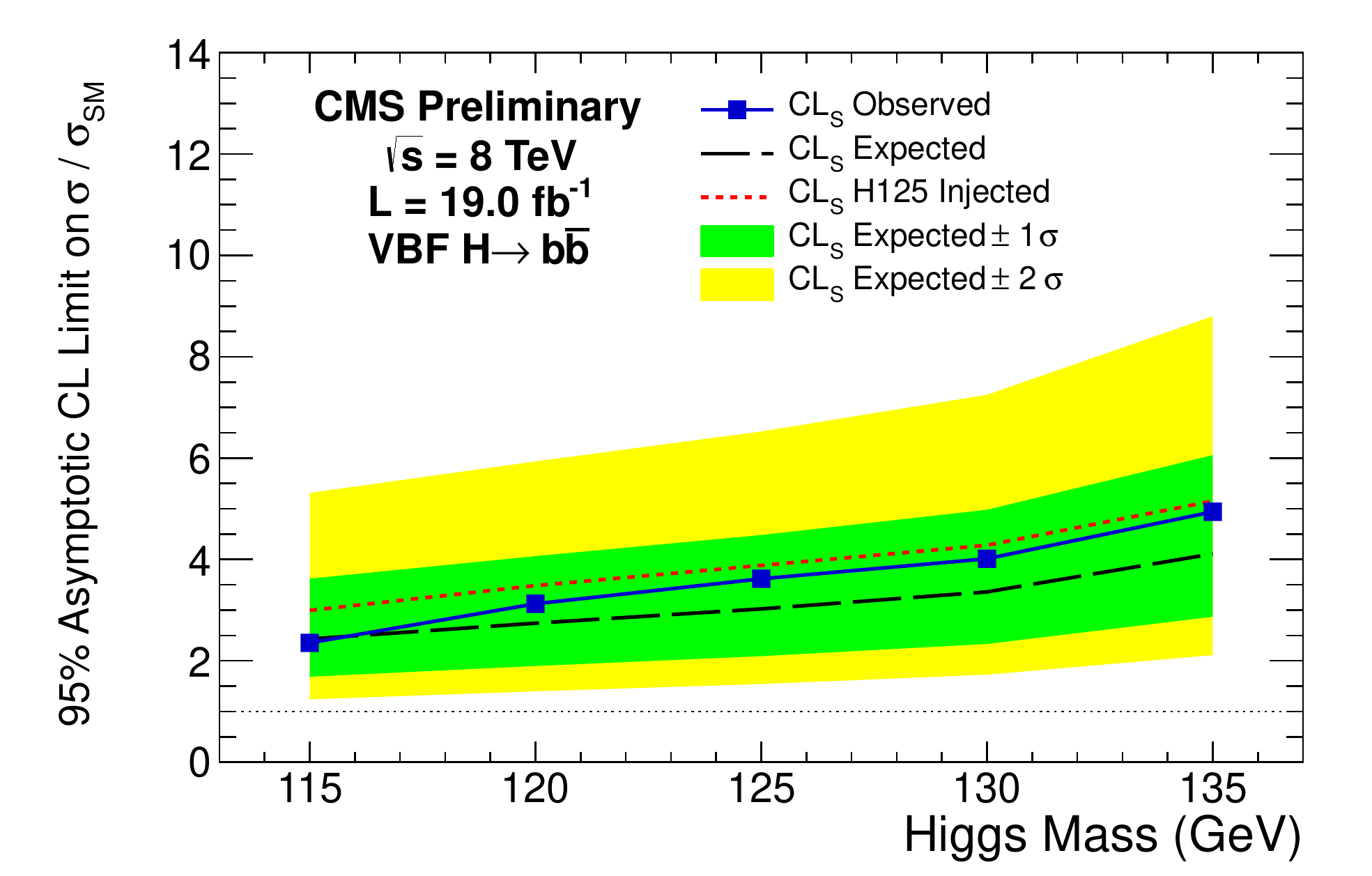}
    \caption{Expected and observed 95\% confidence level limits on the signal cross section in units of the SM
expected cross section.}
     \label{fig:limitsVBF}
  \end{center}
\end{figure}

Upper limits on the production cross section times the \HBB\ branching ratio, with respect to the expectations for a SM Higgs boson, are derived for a Higgs boson in the mass range 115--135 GeV and shown in Fig.~\ref{fig:limitsVBF}. In this range, the expected 95\% confidence level upper limits in the absence of a signal vary from 2.4 to 4.1 times the SM prediction, while the corresponding observed upper limits vary from 2.4 to 5.2. 
At a Higgs boson mass of 125 GeV the expected limit is 3.0 and the observed limit is 3.6. For a 125 GeV Higgs boson signal the observed (expected) significance is  0.5 (0.7)$\sigma$, and the fitted signal strength is $\mu=\sigma/\sigma_{\rm SM}=0.7\pm1.4$. Because of the small signal to background ratio, the results are dominated by the statistical uncertainty in the background.

\section{ $\ttH$ Production }
The $\ttH$ production paired with the \HBB\ decay mode is an interesting process, whose rate depends on the largest of the fermionic couplings to the Higgs boson - top and bottom quarks - which are two key couplings to probe the Higgs boson's consistency with SM expectations.
Although the $\ttH$ production contributes little to the expected cross section with respect to the total Higgs boson production (0.6\%), this signature provides a probe that is complementary to the VH channel: they both provide information about the interaction between the bottom quark and the Higgs boson, but the dominant backgrounds are very different, \ttbar+jets production instead of W+jets production.

The $\ttH$ vertex is the most challenging one to probe directly, but it represents the only way to probe the coupling of the Higgs boson to top quark in a model-independent manner. Since the expected SM rates in this channel are very small, a sizable excess would be clear evidence for new physics~\cite{tth-12-9}.

\begin{figure}[h!]
\centering
 \includegraphics[width=0.67\linewidth]{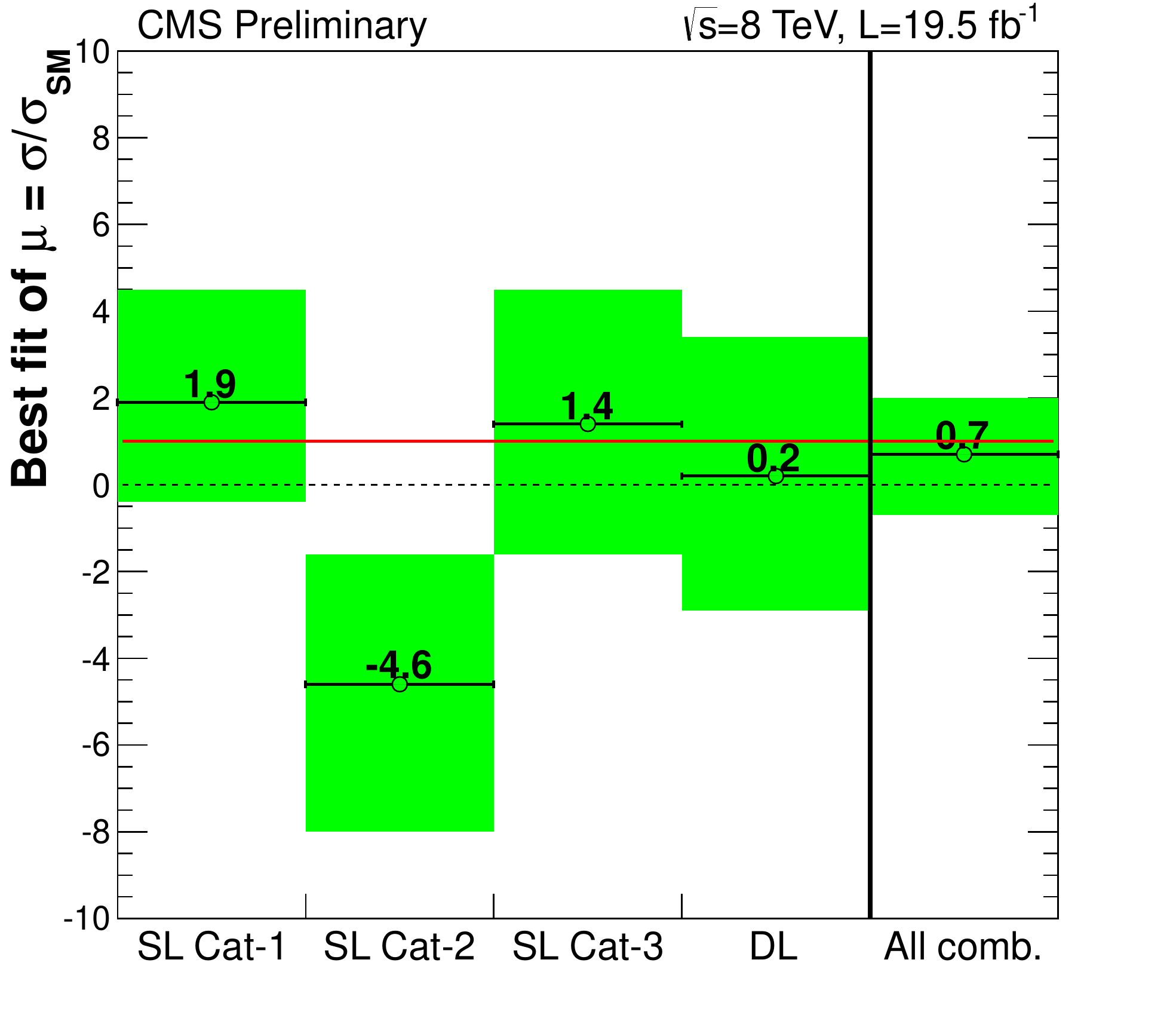}
\caption{The best-fit value of the signal strength modifier $\mu$, broken-up by category~\cite{PAS-14-010}.}
\label{fig:tth2} 
\end{figure}

The final state has a large multi-jet background that is suppressed to a tiny level by requiring at least one charged lepton (electron or muon) in the event.~The resulting signature is still dwarfed by the QCD production of
a \ttbar pair plus additional jets.  Due to the good experimental efficiency to tag jets that originate from b quarks, while rejecting light-flavored hadron decays, the background is greatly reduced by requiring at least two jets in the final state to be b-tagged. Still, the background subprocess $\mathrm{pp}\to\ttBB$ remains irreducible since it has the same final state of the signal. It provides experimental challenges since its cross section is much larger than that
of the signal and its rate and shape have also large theoretical uncertainties~\cite{ttbb}.
%The search for $\ttH$ in the $\HBB$ decay mode using the full 8 TeV data set has been performed by ATLAS,~\cite{PASttHbbATLAS}, reports no significant excess of events above the background expectation and an observed (expected) 95\% confidence-level limit of 4.1 (2.6) times the SM cross section is obtained. The measured signal strength relative to the SM expectation is found to be $\mu = 1.7 \pm 1.4$. 
Recently, CMS has reported new results~\cite{PAS-14-010} on this search which exploits the Matrix Element Method~\cite{MEM1} technique to simultaneously achieve both a mitigation of the combinatorial background and also a maximal separation between the signal and the irreducible $\mathrm{p}\mathrm{p}\to\ttBB$ background. 
This analysis reported as observed limit $\mu<3.3$, corresponding to a best--fit value of ${\mu}=0.67^{+1.35}_{-1.33}$, Fig.~\ref{fig:tth2}.

\section{Summary}
Searches for the standard model Higgs boson decaying to \bbbar exploiting the VH, VBF and ttH production modes are reported. Data collected with the CMS experiment corresponding to
integrated luminosities of up to 5 fb$^{-1}$ at $\sqrt{s}=$7~TeV and up to 20 fb$^{-1}$ at $\sqrt{s}=$8~TeV are analysed. 
The VH channel, as the most sensitive search, reports an excess of events above the expected background with a local significance of 2.1 standard deviations compatible with a Higgs boson mass of 125 GeV. The signal strength corresponding to this excess is ${1.0}_{-0.5}^{+0.5}$. The fitted signal strengths for the VBF and ttH searches are respectively $\mu=0.7\pm1.4$ and ${\mu}=0.67^{+1.35}_{-1.33}$.

%% The Appendices part is started with the command \appendix;
%% appendix sections are then done as normal sections
%% \appendix

%% \section{}
%% \label{}

%% References
%%
%% Following citation commands can be used in the body text:
%% Usage of \cite is as follows:
%%   \cite{key}         ==>>  [#]
%%   \cite[chap. 2]{key} ==>> [#, chap. 2]
%%

%% References with BibTeX database:
\nocite{*}
\bibliographystyle{elsarticle-num}
\bibliography{martin}

%% Authors are advised to use a BibTeX database file for their reference list.
%% The provided style file elsarticle-num.bst formats references in the required Procedia style

%% For references without a BibTeX database:

\end{document}